\documentclass[twocolumn,showpacs,preprintnumbers,prd,superscriptaddress,nofootinbib,floatfix]{revtex4}

\usepackage{amsmath}
\usepackage{amsfonts}
\usepackage{graphicx}
\usepackage{hyperref}

\newcommand{\sla}[1]{\label {#1}}

\begin{document}

\title{Dark matter from dark energy-baryonic matter couplings}

\author{Alejandro Avil\'es}
\email{avilescervantes@gmail.com}
\affiliation{Instituto de Ciencias Nucleares, UNAM, M\'exico}
\affiliation{Depto. de F\'{\i}sica, Instituto Nacional de
Investigaciones Nucleares, M\'exico}
\author{Jorge L. Cervantes-Cota}
\email{jorge.cervantes@inin.gob.mx}
\affiliation{Depto. de F\'{\i}sica, Instituto Nacional de
Investigaciones Nucleares, M\'exico}
\affiliation{Berkeley Center for Cosmological Physics, University of California, Berkeley, California,
USA}

\begin{abstract}
We present a scenario in which a scalar field dark energy is coupled to the trace of the energy momentum tensor of the baryonic matter fields.
In the slow-roll regime, this interaction could  give rise to the cosmological features of dark matter. We work out the cosmological background solutions
and fit the parameters of the model using the Union 2 supernovae data set. Then, we develop cosmological perturbations up to linear order, and we find
that the perturbed variables have an acceptable behavior, in particular the density contrast of baryonic matter grows similar to that in the 
$\Lambda$CDM model for a suitable choice of the strength parameter of the coupling.
\end{abstract}
\date{}
\pacs{98.80.-k, 98.80.Cq, 95.35.+d}
\preprint{}
\maketitle

\section{INTRODUCTION}
Currently, the most successful model of cosmology that we have is $\Lambda$CDM. It is constructed in order to match a wide variety of modern
cosmological observations that have stunned the physicist community in the last decades. Among them, there are precision measurements of  
anisotropies in the cosmological microwave background radiation \cite{WMAP,WMAP7y}, baryon acoustic oscillation \cite{Eisenstein05,Percival07}, 
and Type Ia supernovae \cite{Riess98,Perlmutter99,Union2}. All these observations point out that our Universe at present is  dominated by a 
cosmological constant -dark energy- and there are about 5 times of some unknown 
nonbaryonic, dark matter, over the baryonic matter which is well understood by the standard model of particles.

Nonetheless, some objections to the $\Lambda$CDM model exist, both theoretical \cite{Weinberg,CopSamTsu06} and observational (see for example 
\cite{Peebles10}). Thus, alternative proposals have appeared in the literature giving rise to the idea that dark energy varies with time. Scalar fields 
have attracted special attention, mainly quintessence \cite{RatraPeeb88,Caldwell98,Copeland98}, among some other alternatives \cite{CopSamTsu06}.

Recently there has been a lot of interest in studying couplings in the dark sector species \cite{Amendola99,Bean01,Amendola02,FarPeeb04,Koivisto05,Bean08,
Corasaniti08,Caldera09,Costa09,Mota07}. This is in part motivated by the fact that until today we 
can only extract information of these components through gravitational interaction, a feature that has been dubbed
{\it Dark Degeneracy} \cite{Hu99,Rubano02,Wasserman02,Kunz20091,Kunz20092}. Specifically, we can define the energy content of the dark sector, and 
in fact the dark sector itself, using the Einstein field equations 
\begin{equation}
 8 \pi G T_{\mu\nu}^{dark} = G_{\mu\nu} -  8 \pi G  T_{\mu\nu}^{obs},
\end{equation}
where $G_{\mu\nu}\,$ comes from the observed geometry of the Universe and  $T_{\mu\nu}^{obs}\,$ from its observed energy content. In this sense
the dark sector reflects our lack of knowledge. The easiest way -mathematically and conceptually- to accommodate these ideas into the observed 
history of the Universe is decomposing $T_{\mu\nu}^{dark}$  in two species, dark energy and dark matter,
\begin{equation} \sla{canondecomp}
T_{\mu\nu}^{dark} = T_{\mu\nu}^{DE} + T_{\mu\nu}^{DM}.
\end{equation}
But this decomposition is not unique. In fact, the dark sector could be composed by a large zoo of particles and complicated interactions 
between them. Or, it could be even just one
unknown field. To accomplish this last possibility we note that in the $\Lambda$CDM model the equation of state parameter of the total 
dark sector is given by

\begin{equation} \sla{lcdmweff}
 w_{eff}   \equiv \frac{\Sigma \rho_i w_i}{\Sigma \rho_i }\simeq - \frac{1\,\,}{1+ \,0.315\, a^{-3}},
\end{equation}
where $i$ index the dark matter component and the cosmological constant, and in the last equality we have used 
$\Omega^{(0)}_{DM} / \Omega^{(0)}_{DE} \approx 0.315$ \cite{WMAP7y}. In order to mimic this model with just one dark field, we must have 
$ w_{dark} \simeq  w_{eff}$. Any fluid with equation of state parameter equal to $ w_{eff}$ will produce the same expansion history
of the Universe.

On the other hand, some string theory-inspired models of dark energy share the peculiarity that scalar fields, like the dilaton, couple directly to matter 
with gravitational strength. To have  cosmological influence at present, these fields must be nearly massless, leading to long-range
fifth-forces and to large violations of the equivalence principle. Some mechanisms have been proposed in order to avoid this unacceptable behavior, as the 
Damour-Poliakov effect \cite{Damour90,Damour94}, in which the interaction is dynamically driven to zero by the expansion of the Universe, and as the 
chameleon mechanism \cite{Khoury041,Khoury042}, where the mass of the scalar field has an environment density dependence, becoming very huge in overall 
high density regions, such as those in which Einstein principle equivalence and fifth-force search experiments are performed.

In this work we follow the both lines of thought outlined above. To this end we consider an interaction Lagrangian between a nearly massless scalar 
and the trace of the energy momentum tensor of the ordinary matter fields given by

\begin{equation} \sla{intlag}
 \mathcal{L}_{int} = \sqrt{-g} A(\phi) T.
\end{equation}

This type of coupling has been investigated in a cosmological context in \cite{NarPad85,SinghPad88,SamiPad03}. This interaction
has some attractive properties, the field does not couple to the electromagnetic field  and in this sense is dark; also, it does not couple 
to relativistic matter and then do not affects the success of the early Universe cosmology, although it could couple to the inflaton field. 
As we shall see, the coupling could also give mass to the field.

At a more fundamental level we can consider, in a first approximation, a fermionic matter free fields with energy momentum tensor 
$T_{\mu\nu}^{(f)} = -i \bar{\psi} \gamma_{\mu}\partial_{\nu} \psi\,$ and trace given by 
$T^{(f)} =  -i \bar{\psi} \gamma^{\mu}\partial_{\mu} \psi = - m_{\psi} \bar{\psi}\psi$, where in the last equality we have used 
the Dirac equation. The coupling then becomes a Yukawa-like one and gives mass to the fermions. If we chose correctly the function
$A(\phi)$ we can also interpret this result as that the interaction has given mass to the field $\phi$.
 
This paper is organized as follows: in Section II, we present the details of the general theory; in Section III, we derive the background cosmology
equations and we choose a specific model that mimics the $\Lambda$CDM model, then we numerically obtain the cosmological solutions; in Section 
IV, we work the theory of linear perturbations; finally, in Section V, we present our conclusions.

\section{The General Theory}

The action we consider is 

\begin{eqnarray} \sla{genaction}
 S &=& \int d^4 x \sqrt{-g} \left[ \frac{R}{16 \pi G} - \frac{1}{2} \phi^{,\alpha} \phi_{,\alpha} - V(\phi)\right] 
\nonumber\\ & &+ S_{int} + S_{m},
\end{eqnarray}
where the coupling of the scalar field with the trace of the energy momentum tensor of the ordinary matter fields is given by  (\ref{intlag}).
There is a difficulty here because the matter fields not only appears in the matter action, but also in the interaction action. Thus, we have to define 
the trace of the energy momentum tensor of baryonic matter as 

\begin{equation} \sla{tracedef}
 T = - \frac{2}{\sqrt{-g}} \frac{\delta (S_{int} + S_{m})}{\delta g^{\mu\nu}} g^{\mu\nu}.
\end{equation}

To solve this redundant definition  we will be specific and we will work with a perfect fluid of dust in the matter sector, the Lagrangian
is $\mathcal{L}_m = - \rho \sqrt{-g}$, where $\rho$ is the energy density of the fluid in its rest frame \cite{deFelice}. 
The trace of the energy momentum tensor becomes (see Appendix, or for an alternative derivation see \cite{SamiPad03})

\begin{equation} \sla{trace1}
 T = -\frac{1}{1-A(\phi)} \rho. 
\end{equation}

Then, the total action can be written  as

\begin{equation} \sla{act2}
 S = \int d^4 x \sqrt{-g} \left[ \frac{R}{16 \pi G} - \frac{1}{2} \phi^{,\alpha} \phi_{,\alpha} 
-V(\phi) - e^{\alpha(\phi)} \rho \right],
\end{equation}
where we have defined the function $e^{\alpha(\phi)}$ through 

\begin{equation}
e^{\alpha(\phi)} = \frac{1}{1-A(\phi)}.
\end{equation}

These theories reduce to those with an interaction term $f(\phi) \mathcal{L}_m$, proposed in \cite{Damour90}, in the case in which the matter fields are
perfect fluids, although the interaction vanishes for relativistic fluids (see Appendix). The field equations derived from action \ref{act2} are

\begin{equation} \sla{EE}
 G_{\mu\nu} = 8 \pi G ( T_{\mu\nu}^{(\phi)} + e^{\alpha(\phi)}T_{\mu\nu}^{(m)} ),
\end{equation}
where the energy momentum tensors of the fields are
\begin{equation} \sla{EMTphi}
T_{\mu\nu}^{(\phi)} =  \phi_{,\mu}\phi_{,\nu}- \frac{1}{2} g_{\mu\nu} \phi^{,\alpha} \phi_{,\alpha}- g_{\mu\nu} V(\phi),
\end{equation}
and
\begin{equation} \sla{EMTdust}
T_{\mu\nu}^{(m)} = \rho \, u_{\mu}u_{\nu}.
\end{equation}  

For the scalar field the evolution equation is 
\begin{equation} \sla{efe}
  \square \,\phi - V'(\phi) =   \alpha'(\phi) e^{\alpha(\phi)} \rho,
\end{equation}
where prime means derivative with respect to $\phi$.
This equation shows that, as in the chameleon models, the evolution of the scalar field is governed by the effective potential $V_{eff}(\phi) = V(\phi) 
+ e^{\alpha(\phi)}  \rho$, and that the mass of slow oscillations about the minimum is given by $m^2 = V''_{eff}(\phi_{min})$, where $\phi_{min}$ 
is the value of the scalar field that minimizes the effective potential. By using the Bianchi identities the conservation equation for the matter fields 
becomes

\begin{equation} \sla{consm} 
 \nabla^{\mu} T_{\mu\nu}^{(m)} = - \alpha'(\phi) \rho(g_{\mu\nu} + u_{\mu}u_{\nu}) \partial^{\mu} \phi.
\end{equation}
 
From this equation we obtain the continuity and geodesic equations

\begin{equation} \sla{conteq}
 \nabla^{\mu} ( \rho \, u_{\mu}) = 0,
\end{equation}
and
\begin{equation} \sla{geodeq}
 u^{\mu}\nabla_{\mu} u^{\nu} = -   \alpha'(\phi) (g^{\mu\nu} + u^{\mu}u^{\nu})  \partial_{\mu} \phi.
\end{equation}

The geodesic equation (\ref{geodeq}) shows that only the chan\-ges of the scalar field perpendicular 
to the four velocity of a particle affect its motion. In particular, if the vector field $u^{\mu}$ defines a global time, only the spatial
gradient of the field influences the particle's motions. This is the case of the homogeneous and isotropic cosmology, in which there is a family of 
preferred comoving observers  with the expansion. If the field is also homogeneous and isotropic, these observers are affected  by
the field only through gravity.

In the nonrelativistic limit with static fields, the geodesic equation (\ref{geodeq}) for a point particle reduces to

\begin{equation} \sla{WFlim}
 \frac{d^2 \vec{X}}{d t^2} = - \nabla \Phi_N - \alpha' \nabla \phi -  \alpha' \vec{V} (\vec{V} \cdot \nabla)\phi,
\end{equation}
where $\vec{X}$ is the spatial position of the particle, $\vec{V}$ its three velocity and $\Phi_N = -\frac{1}{2} h_{00}$ the Newtonian potential.
The last term can be dropped out because it is of the order $(v/c)^2$, but we prefer to leave it there because it is a dissipative term that arises due 
to the interaction with the scalar field and it shows explicitly how the energy transfer occurs. Equation (\ref{WFlim}) also shows that the 
force exerted by the scalar field upon a test particle of mass $m$ is given by 

\begin{equation} \sla{phiforce}
 \vec{F}_{\phi} = - m \nabla ( \alpha(\phi) ).
\end{equation}

\section{Background Cosmology}

Considering a  spatially flat FRW metric, we write the background evolution equations as

\begin{eqnarray} 
 H^2 = \frac{8 \pi G}{3} \Big(\frac{1}{2} \dot{\phi}^2 + V(\phi) \!\!&+&\!\! (e^{\alpha(\phi)}-1) \rho + \rho \Big), \label{F1a}\\
\ddot{\phi} + 3 H \dot{\phi} + V'(\phi) \!\!&+&\!\!  \alpha'(\phi) e^{\alpha(\phi)} \rho = 0, \label{F1b}\\
\dot{\rho} + 3 H \rho \!\!&=&\!\! 0. \label{F1c}
\end{eqnarray}
The last equation is a consequence of (\ref{consm}) (see also the discussion after equation (\ref{geodeq})), and it can be integrated  
to give $\rho=\rho_0 a^{-3}$, where $\rho_0$ is the value of the 
energy density of baryonic matter today and we have normalized the value of the scale factor today to $a_0 =1$. This set of equations is a general
feature of interactions of the type $\,f(\phi)\mathcal{L}_m = -f(\phi) \rho$ (see for instance \cite{Bean01,Koivisto05,Bean08}). The conservation equation
suggests to interpret the interaction as a pure dark sector feature. Then we define

\begin{equation}
 \rho_{dark} = \rho_{\phi} + (e^{\alpha(\phi)} -1) \rho,
\end{equation}
where  $\rho_{\phi} = \dot{\phi}^2 /2 + V(\phi)$. The total energy density is then $\rho_T = \rho_{dark} + \rho$, and the 
Friedman equation, $3  H^2 = 8 \pi G (\rho_{dark} + \rho )$. The conservation equation for  $\rho_{dark}$ is

\begin{equation}
 \dot{\rho}_{dark} + 3 H (1 + w_{dark}) \rho_{dark} = 0,
\end{equation}
where

\begin{equation}
 w_{dark} = \frac{ \dot{\phi}^2/ 2 -V (\phi)}{\dot{\phi}^2/ 2 + V (\phi) +
(e^{\alpha(\phi)}-1) \rho_0 a^{-3}}.
\end{equation}

If the field is in  slow-roll regime we neglect the $\dot{\phi}^2$ contribution and the dark equation of state parameter becomes

\begin{equation}
 w_{dark} = -\frac{1}{1+ \frac{e^{\alpha(\phi)}-1}{V (\phi)}\rho_0 a^{-3}}.
\end{equation}

Then, the following cosmological scenario is plausible without dark matter. During the radiation era, the trace of the energy momentum tensor of 
the baryonic matter is equal to zero and the background cosmology is  as the standard big bang. Then, in the early times of matter era  
the energy density $\rho$ is  high, and thus the interaction term dominates in the last 
expression, giving rise to $w_{dark} \approx 0$. As the time goes on, $\rho$ is redshifted as $\sim a^{-3}$,  the potential 
becomes important and  eventually dominates over the interaction term,  and in this limit $ w_{dark} \rightarrow -1$.

To obtain an acceptable late time cosmology we have to impose a set of constrictions in the free functions of the theory. First, 
to obtain equation (\ref{lcdmweff}) we impose the constriction 

\begin{equation} \sla{1fc}
C_1{(\phi)} \equiv \frac{e^{\alpha(\phi)}-1}{V (\phi)}\rho_0 \simeq \frac{\Omega^{(0)}_{DM}}{\Omega^{(0)}_{DE}}.
\end{equation}

Second, to account for the correct amount of {\it dark matter}, here given by the interaction term, we impose the constriction

\begin{equation} \sla{2fc}
C_2{(\phi)} \equiv e^{\alpha(\phi)}-1 \simeq \frac{\Omega^{(0)}_{DM}}{\Omega^{(0)}_{b}}.
\end{equation}  

$C_1$ and $C_2$ are general functions of the scalar field that we expect to have a very little variation in the slow-roll regime. 
Also, to obtain the observed accelerated expansion of the Universe, we must have
\begin{equation} \sla{3fc}
 8 \pi G V(\phi_0) = \Lambda,
\end{equation}
where $\phi_0$ is the value of the scalar field today and $\Lambda$ is the measured {\it cosmological constant}. Note that this last 
equation is not independent from the two constrictions.

The interaction over fermions impose other constraints because, as we show in the Appendix, its effect 
is to shift the fermion masses from $m\,$ to $m(\phi)= e^{\alpha(\phi)}m$.   Following \cite{Brax04} the abundance of light elements 
constrains the variation of protons' and neutrons' masses to be at most of about $10\%$ from nucleosynthesis until today.  In our 
model this is translated into  $\Delta m(\phi)/m(\phi) = (e^{\alpha(\phi_n)} - e^{\alpha(\phi_0)})/ e^{\alpha(\phi_0)} <  0.1$, 
where $\phi_n$ is the value of the scalar field at nucleosynthesis and $\phi_0$ is the value of 
the scalar field today.  According to this the scalar field has to be in slow-roll at least since that epoch.

\subsection*{Fixing the free functions}

In the chameleon model the potential is of the runaway form and the coupling with matter is an exponential one. The effective equation of 
state parameter becomes of the form $w_{dark} = -(1+ \phi(\exp(\beta \phi)-1)a^{-3})^{-1}$ and the arguments shown above are applicable as long 
as the field is in the slow-roll regime. Nevertheless we will follow a different approach and with the only aim to mimic as far as possible the 
$\Lambda$CDM model we choose the free functions of the theory as

\begin{equation} \sla{ft1}
e^{\alpha(\phi)}= 1 + \frac{1}{2} \epsilon \phi^2\,,
\end{equation}
and
\begin{equation} \sla{ft2}
V(\phi) = \frac{1}{2}m^2_{\phi} \phi^2,
\end{equation}
where $\epsilon$ is a constant with inverse squared mass units. Note that with this selection the minimum of the effective potential is always at $\phi=0$,
and the evolution field equation for the scalar field (\ref{efe}) becomes

\begin{equation} \sla{efe2}
  \square \,\phi - (m^2_{\phi} + \epsilon \rho) \phi =  0,
\end{equation}
and  it is explicit that the field has an effective environment dependent  mass 
\begin{equation} \sla{effmass}
 m^2_{eff} = m^2_{\phi} + \epsilon \rho,
\end{equation}
and that the force mediated by the field has a range $1/m_{eff}$. Although this is a feature on chameleon fields, the evolution equation here is 
linear and then we do not expect to obtain the thin-shell suppression \cite{Khoury041,Khoury042}, also a feature in chameleon fields.

The two constriction equations (\ref{1fc}) and (\ref{2fc}) become

\begin{equation} \sla{1nc}
C_1 = \frac{\epsilon \rho_0}{m^2_{\phi}} \simeq \frac{\Omega^{(0)}_{DM}}{\Omega^{(0)}_{DE}} \simeq 0.315
\end{equation}
and
\begin{equation} \sla{2nc}
C_2(\phi) = \frac{1}{2}\epsilon \phi^2 \simeq \frac{\Omega^{(0)}_{DM}}{\Omega^{(0)}_{b}} \simeq 5.
\end{equation}

The equation that relates the scalar field mass with $\Lambda$ becomes
\begin{equation} \sla{3nc}
m^2_{\phi} \simeq 2 \Lambda \left(\frac{M_p}{\phi_{0}} \right)^2,
\end{equation}
where the reduced Planck mass is given by $M_p= (8 \pi G)^{-1/2}$.

\subsection*{Numerical Analysis}

With the selection of the free functions given in  last section, the cosmological background evolution equations can be written as

\begin{equation} \sla{F2}
 H^2 = \frac{8 \pi G}{3} \Big(\frac{1}{2} \dot{\phi}^2 + \frac{1}{2} m^2_{\phi} \phi^2 + \frac{1}{2} \epsilon \phi^2 \frac{\rho_0}{a^3} 
 + \frac{\rho_0}{a^3} \Big),
\end{equation}
and 
\begin{equation} \sla{F3}
\ddot{\phi} + 3 H \dot{\phi} + m^2_{\phi} \phi  + \epsilon \frac{\rho_0}{a^3} \phi  = 0.
\end{equation}

To do the numerical analysis, it is convenient to define the dimensionless strength parameter of the interaction
\begin{equation}
 \beta \equiv \epsilon M^2_p.
\end{equation}

Using $H_0 = 70.4 Km/s/Mpc$, $\Lambda/3 H_0^2 =\Omega_{\Lambda} = 0.73$ and equation (\ref{3nc}), the  required value of the scalar field mass 
as a function of $\beta$ and $C_2(\phi_0)$ is 
\begin{equation} \sla{mosf}
 m_\phi = 3.6 \times 10^{-4} \sqrt{\frac{\beta}{C_2(\phi_0)}}\, \text{Mpc}^{-1}. 
\end{equation}

We evolve the equations (\ref{F2}) and (\ref{F3})  for different values of $\beta$, using the numerical constrictions 
(\ref{1nc}) and (\ref{2nc}).  

\begin{figure}[ht]
\centering
\includegraphics[width=2.2in]{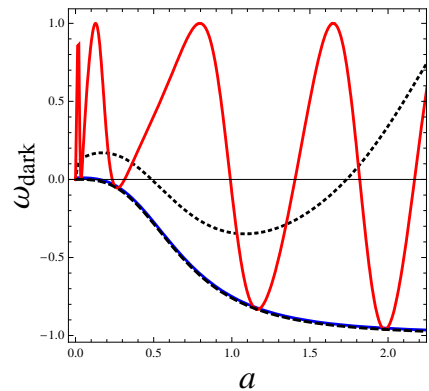}   
\caption{Shows the evolution of the equation of state parameter of the dark sector, $w_{\text{dark}}$. 
The dashed line is the result in the $\Lambda$CDM model. The thick oscillating curve  corresponds to $\beta=  1$; the short-dashed, 
to $\beta = 0.2$; and the thick nonoscillating to $\beta = 0.04$.}
\label{fig1}
\end{figure}

\begin{figure}
\centering
\includegraphics[width=2in]{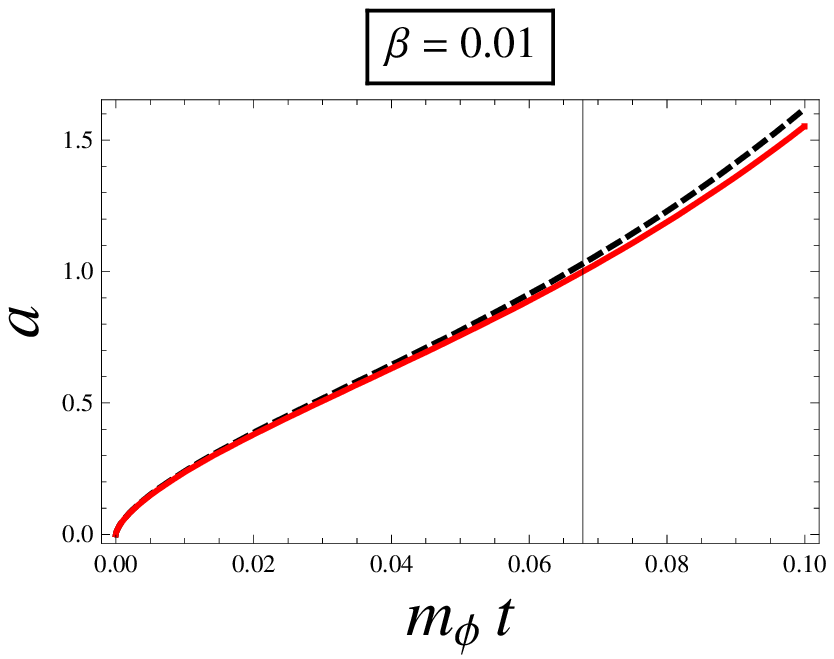}
\includegraphics[width=2in]{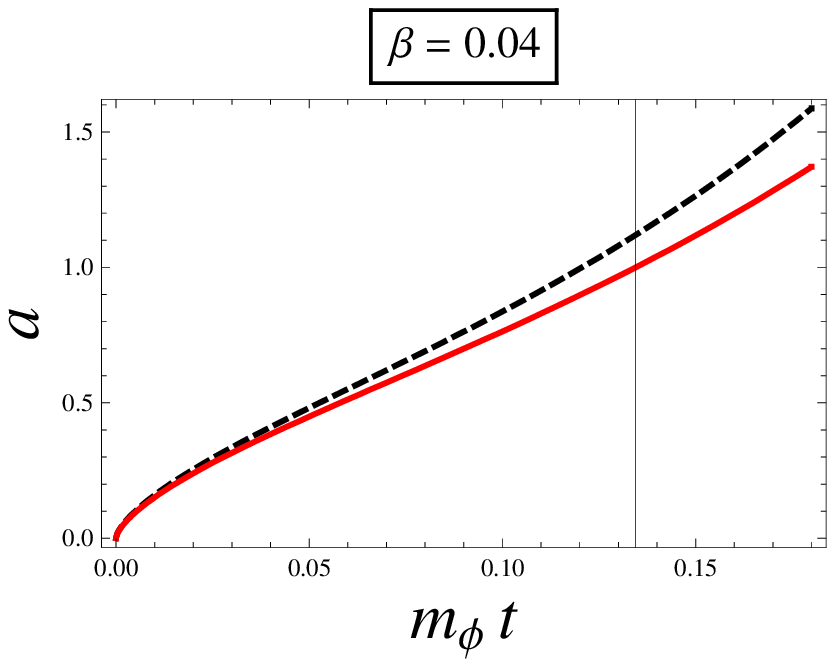}     
\caption{Evolution of the scale factor as a function of cosmic time. The dashed lines are the results for the $\Lambda$CDM model. The vertical lines 
denote the present time. Note that the temporal scale is different for each case.}
\label{fig2}
\end{figure}

Figure \ref{fig1} shows the results for $w_{dark}$. The equation of state that is oscillating comes from a field that is oscillating about the minimum at 
$\phi=0$ and then it is not in the slow-roll regime. The nonoscillating curves come from fields that are in the slow-roll regime. Note that the curve 
with $\beta = 0.04$, which corresponds to $\,\epsilon \approx G\,$, comes from a field in slow-roll regime. 

We plot the scale factor $a(t)$ as a function of time. The results are shown in  figure \ref{fig2}. We also show the results for the 
$\Lambda$CDM model, and the initial conditions are such that at some early time both models have the same amount of {\it dark matter}, here 
given by the interaction. Note that the scale factor grows more slowly than in the $\Lambda$CDM model. This is because the emulated dark matter 
component does not redshift with $a^{-3}$, but rather with $\phi^2 a^{-3}$. The age of the Universe is, in the case of $\beta=0.01$, 
about $3 \% $ larger than in the $\Lambda$CDM model, and about $10 \%$ for the case $\beta = 0.04$.
\begin{figure}
\centering
\includegraphics[width=2.1in]{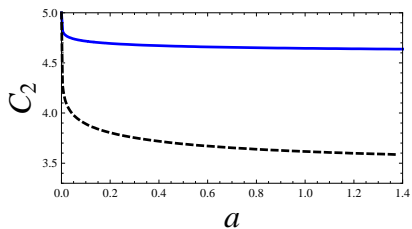} 
\caption{Evolution of  $C_2(\phi)$ as a function of the scale factor using: $\beta=0.01$, thick line;
and $\beta = 0.04$, dashed line.}
\label{fig3}
\end{figure}

\begin{figure}
\centering
  \includegraphics[width=3in]{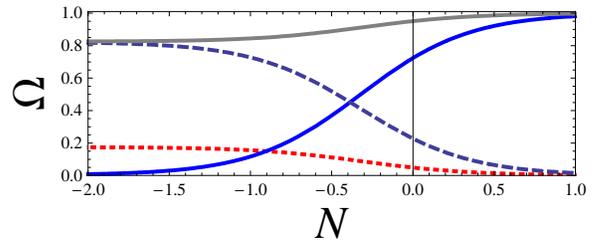} 
\caption{Evolution of the different density parameters as a function of $N=\log a$, we use $\beta=0.01$. 
The upper line is $\Omega_{\text{dark}}$; the dotted line is $\Omega_{b}$; the dashed line, 
$\Omega_{\text{int}}$; and the thick line, $\Omega_{\phi}$. The vertical line denotes present time.}
\label{fig4}
\end{figure}

Figure \ref{fig3} is a plot of the function $C_2(\phi)$ as a function of the scale factor for the strength parameters $\beta=0.01$ and $\beta = 0.04$, with 
initial value $C_2(\phi_i)=5$. Note that in both cases, after an initial transient period, the functions become less steep.

Figure \ref{fig4} shows the evolution of the different density parameters as a function of the number of $e$-folds, $N= \log a$. These are
given by
\begin{eqnarray}
\Omega_{\text{dark}} &=& \frac{8 \pi G}{3 H^2} \left(\frac{1}{2} \dot{\phi}^2 +  \frac{1}{2}  m^2_{\phi} \phi^2 + \frac{1}{2} \epsilon 
                     \phi^2 \frac{\rho_0}{a^3} \right),\nonumber\\
\Omega_{b} &=&  \frac{8 \pi G}{3 H^2} \frac{\rho_0}{a^3}\,,\nonumber\\
\Omega_{\text{int}} &=&  \frac{8 \pi G}{3 H^2} \frac{1}{2} \epsilon  \phi^2 \frac{\rho_0}{a^3},\nonumber\\
\Omega_{\phi} &=& \frac{8 \pi G}{3 H^2} \left( \frac{1}{2} \dot{\phi}^2 +  \frac{1}{2}  m^2_{\phi} \phi^2\right). \nonumber
\end{eqnarray}
Note that $\,\Omega_{\text{dark}} = \Omega_{\text{int}} + \Omega_{\phi}\,$ and $\,\Omega_{\text{dark}} + \Omega_{b} = 1$.

\subsection*{Fit to supernovae data}

\begin{figure}
\centering
  \includegraphics[width=3in]{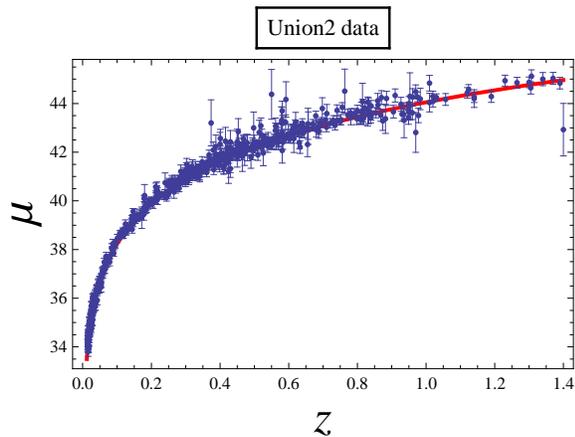} 
\caption{Supernovae Union 2 data and the prediction for the model (solid line). We have used $\beta = 0.01$, 
$C_1 = 0.305$ and $C_2(\phi_0)=4.66$.}
\label{fig4_1}
\end{figure}

Figure \ref{fig4_1} shows the best fit to the Union 2 data set \cite{Union2}, a recent sample of 557 supernovae. We use $\beta=0.01$ and the 
minimum value  of the $\,\chi^2$ function is given by $C_1 = 0.305$ and $C_2(\phi_0)=4.66$, where $\phi_0$ is the cosmological value of the scalar 
field today. 
The module distance $\mu$ and the luminosity distance $d_L$ are related by
\begin{equation}
 \mu = 5 \log_{10} \left(\frac{d_L}{\text{Mpc}} \right) + 25.
\end{equation}

\begin{figure}
\centering
  \includegraphics[width=2.5in]{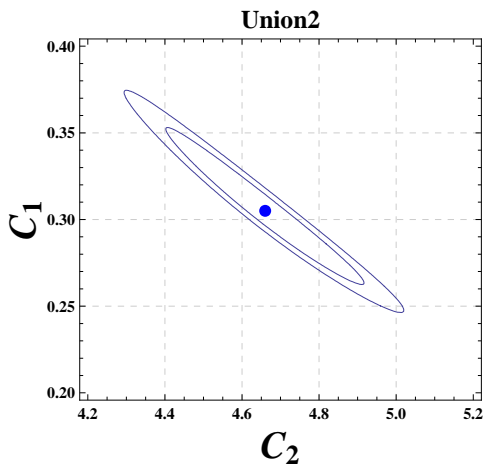} 
\caption{Contour plots at $1\sigma$ and $2\sigma$. The central point is $C_1 = 0.305$ and $C_2(\phi_0)=4.66$, which corresponds to a  
minimum value $\chi_{\rm d.o.f.}^2=0.98$.}
\label{fig4_2}
\end{figure}

Figure \ref{fig4_2} shows the contour plots at $1\sigma$ and $2\sigma$ for this same fit.

In the background cosmological solutions the model looks very similar  as the  $\Lambda$CDM one as far as $\beta \lesssim 0.04$. In the next section we consider the linear perturbation theory of the cosmological solutions.

\section{Linear Perturbations}
In the last section we have shown how the background dynamics can be made very similar to the $\Lambda$CDM model. This 
is mainly because the background fluids are affected by the scalar field only gravitationally, see equation (\ref{F1c}). This
comes ultimately from the fact that the right-hand side of the conservation equation (\ref{consm}) projects out temporal 
variations of $\,\phi$, with respect to the observer $u^{\mu}$. In the background cosmology only temporal variations 
of the scalar field are considered, and so it does not exert any net force over the baryons. 
This is not true in the inhomogeneous cosmology where spatial gradients of the fields have to be taken into 
account. Accordingly one expects that the evolution of cosmological perturbations of baryonic fields could differ 
from the obtained in the $\Lambda$CDM model. In this section we show that the model has an acceptable 
growth of baryonic matter density perturbations.

\begin{figure}
\centering
\includegraphics[width=2.9in]{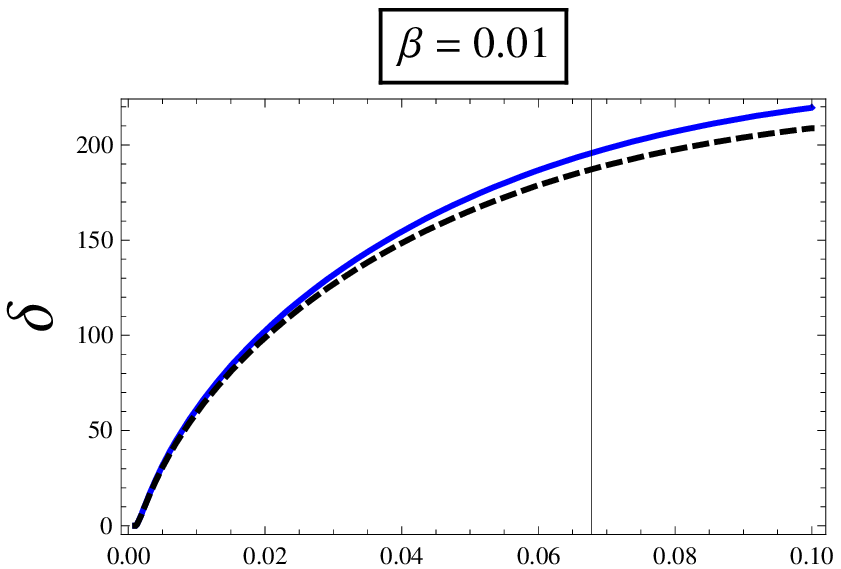}
\includegraphics[width=2.9in]{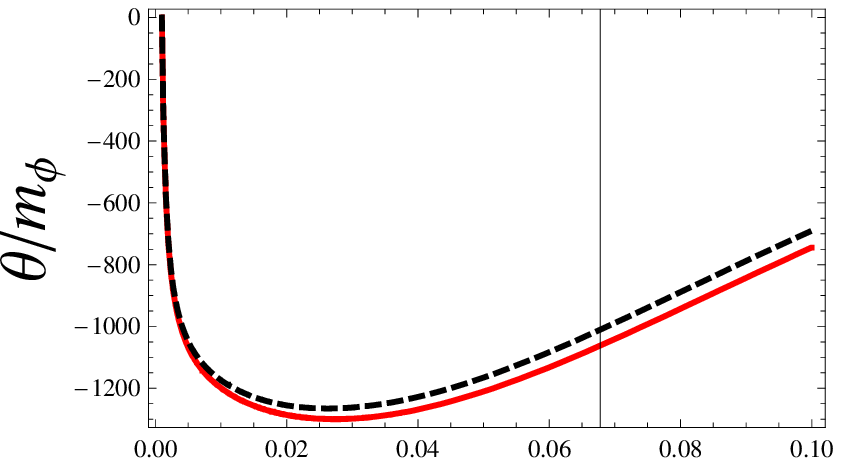}
\includegraphics[width=2.9in]{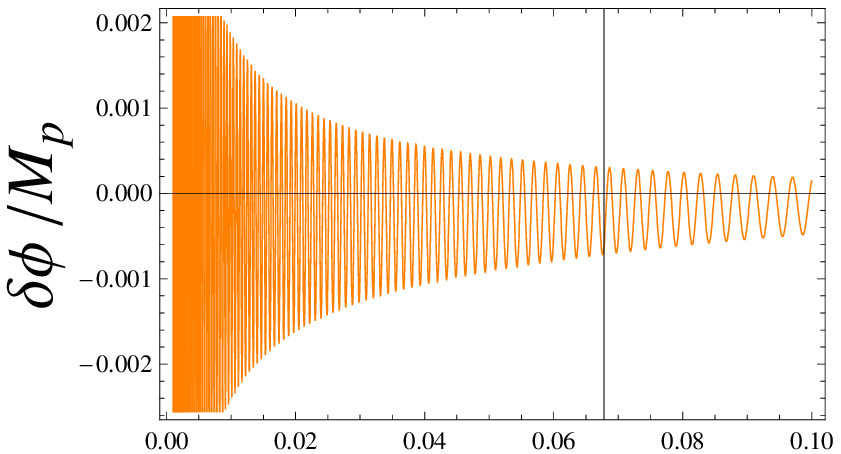}
\includegraphics[width=2.9in]{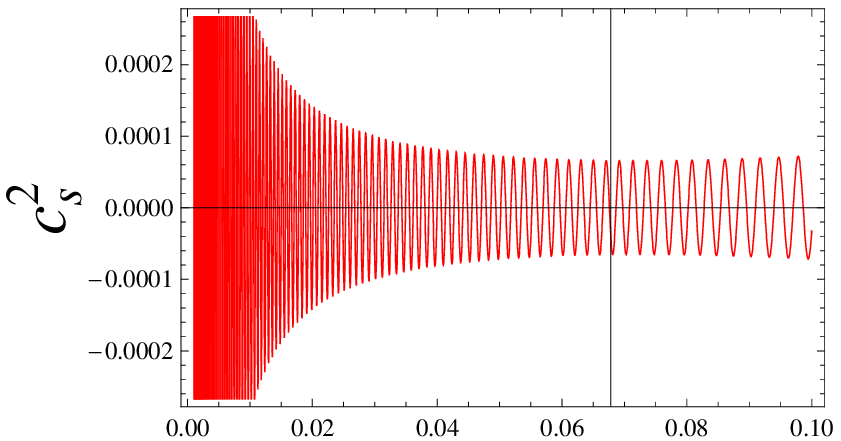}  
\includegraphics[width=2.9in]{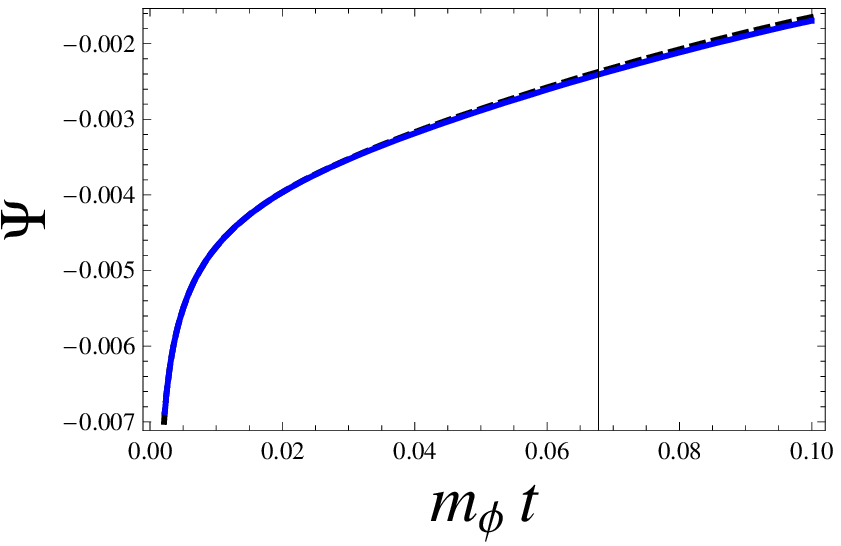}   
\caption{Evolution of the initial perturbations for a comoving wavenumber $k=0.05 \text{Mpc}^{-1}$, using $\beta =0.01$.  
For subhorizon scales $k \gg H$, the relation  $\delta(k_2)/ \delta(k_1) =  k_2^{\,2}/k_1^{\,2}$ for the density contrast holds approximately. 
The dashed lines are the results in the $\Lambda$CDM model and the vertical lines denote present time. The background dynamics are the same as in 
figure \ref{fig2}.}  
\label{fig5}
\end{figure}
\begin{figure}
\centering
\includegraphics[width=2.9in]{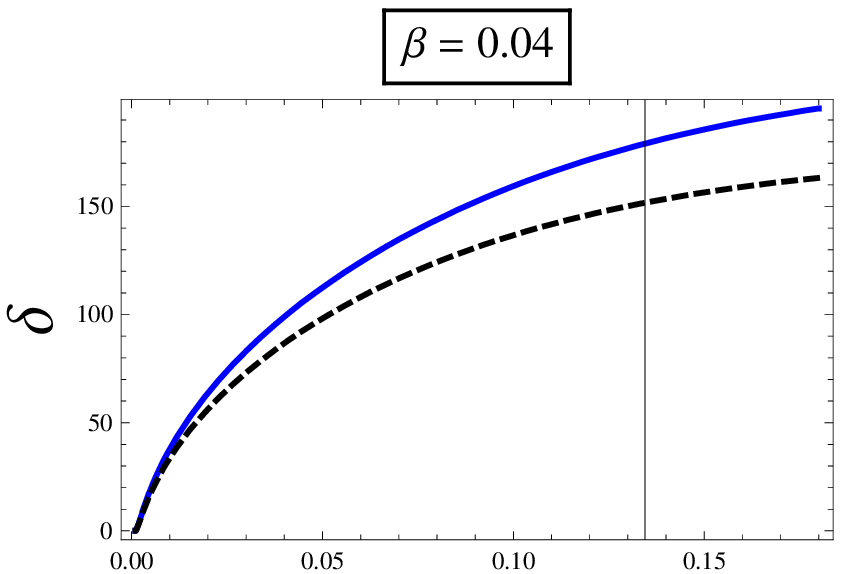}
\includegraphics[width=2.9in]{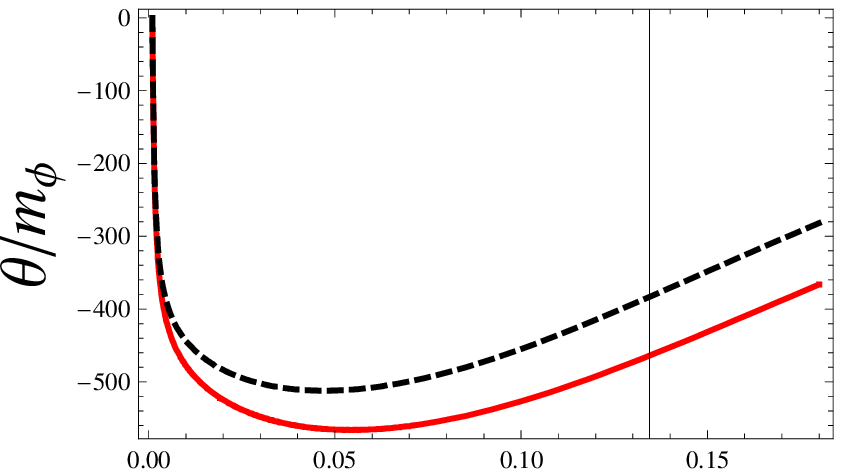}
\includegraphics[width=2.9in]{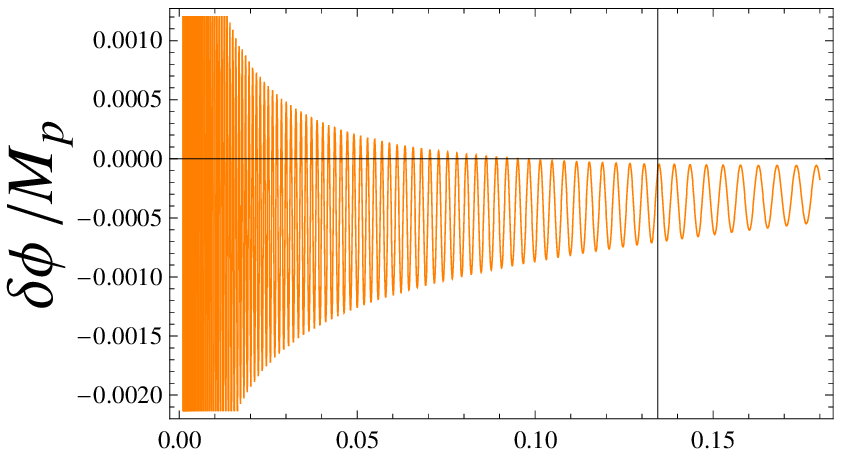}
\includegraphics[width=2.9in]{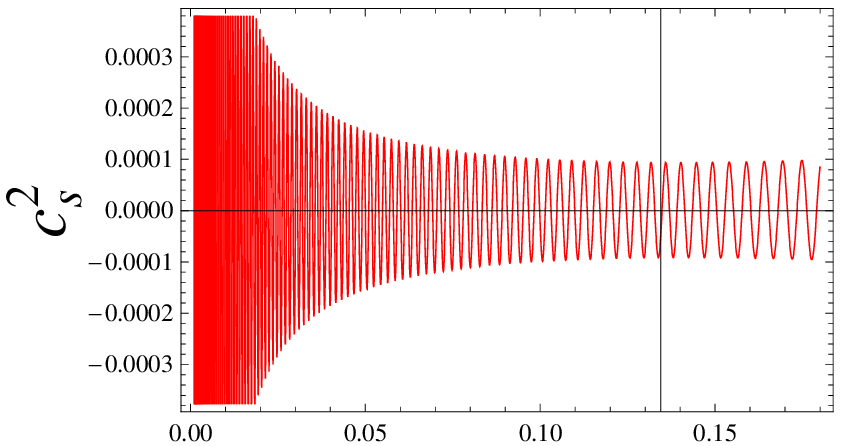}  
\includegraphics[width=2.9in]{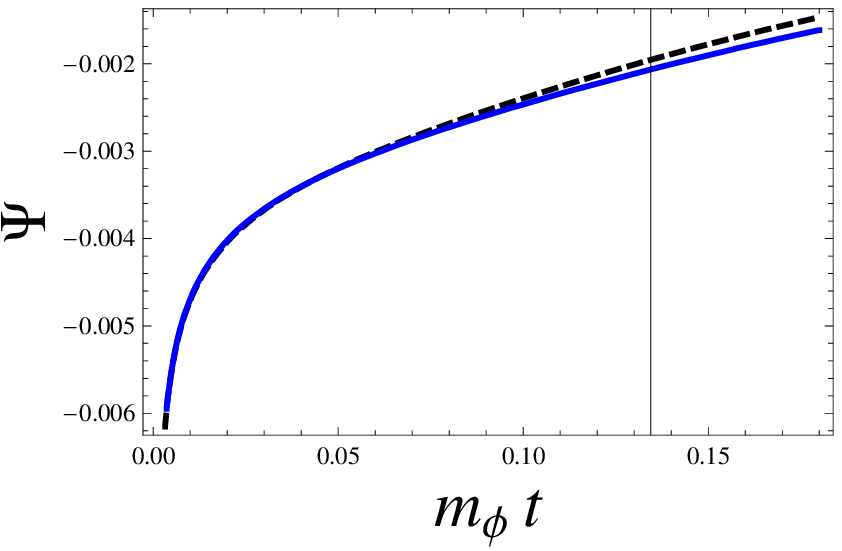}   
\caption{The same as figure \ref{fig5} but with $\beta =0.04$, which corresponds to $\epsilon \approx G$.}  
\label{fig6}
\end{figure}

We consider  scalar perturbations in the longitudinal gauge. The metric is given by
\begin{equation}
ds^2 =  -(1+2\Psi) dt^2 + a^2(t)(1-2\Phi)\delta_{ij} dx^i dx^j. 
\end{equation}

The fields perturbations are given by
\begin{eqnarray} \sla{bmp}
\rho_b(t,\vec{x}) &=& \rho(t)(1+\delta(t,\vec{x})), \\
 \phi(t,\vec{x}) &=&  \phi_0(t) +\delta \phi(t,\vec{x}), \\
 u^{\mu} &=& u_0^{\,\mu} + u_p^{\,\mu},
\end{eqnarray}
where $\rho(t)$  and $u_0^{\,\mu}$ are the background  energy density and velocity of the baryonic matter fluid, 
and $\phi_0(t)$ is the background  value of the scalar  field. As usual we define the variable $\theta$ of the baryonic fluid as the divergence of 
the peculiar velocity of the fluid in momentum space (see \cite{Ma04} for details),
\begin{equation}
 \theta = i k_i v^i, 
\end{equation}
where $k_i$ is the $i$ component of the comoving wave number vector $\vec{k}$ of a perturbation, related to the physical 
wavelength by $\lambda = 2\pi a /k$.

In this analysis we neglect anisotropic stresses and then the scalar potentials are equal, $\Psi=\Phi$. 
The perturbation equations in Fourier space are given by

\begin{eqnarray} \sla{pe1}
 \frac{k^2}{a^2} \Psi \!\!\!\!\!\!&\!& + 3 H(\dot{\Psi} + H \Psi) = 4 \pi G \Big[ \,\dot{\phi}_0^{\,2}\Psi - \dot{\phi}_0 \dot{\delta \phi} 
-m_{\phi}^2\phi_0 \delta \phi \nonumber\\    & &  
- \rho \left( 1+ \frac{1}{2} \epsilon \phi_0^{\,2}\right) 
\Big( \delta  + \frac{\epsilon \phi_0}{1+ \frac{1}{2} \epsilon \phi_0^{\,2}} \delta\phi \Big) \Big],
\end{eqnarray}

\begin{equation} \sla{pe2}
  \frac{2 k^2}{a^2} (\dot{\Psi} + H \Psi) = 8 \pi G \left[\frac{k^2}{a^2}\dot{\phi}_0 \delta \phi + 
\rho \left( 1+ \frac{1}{2} \epsilon \phi_0^{\,2}\right) \frac{\theta}{a} \right],
\end{equation}

\begin{eqnarray} \sla{pe3}
 \ddot{\Psi} + 4 H \dot{\Psi} \!\!\!\!\!\!&\!&+ (2\dot{H} + 3 H^2) \Psi =\nonumber\\
\!&\!&  4 \pi G [ \,-\dot{\phi}_0^{\,2}\Psi + \dot{\phi}_0 \dot{\delta \phi} 
-m_{\phi}^2\phi \delta \phi ],
\end{eqnarray}
and the hydrodynamical perturbation equations

\begin{equation} \sla{pe4}
 \dot{\delta} - 3 \dot{\Psi} + \frac{1}{a}\theta = 0,
\end{equation}

\begin{equation} \sla{pe5}
 \frac{1}{a}\dot{\theta} + \left( 2 H + \frac{\epsilon \phi_0 \dot{\phi}_0}{1+ \frac{1}{2} \epsilon \phi_0^{\,2}} \right) \frac{1}{a}\theta
= \frac{k^2}{a^2}\left(\Psi + \frac{\epsilon \phi_0 }{1+ \frac{1}{2} \epsilon \phi_0^{\,2}}\delta\phi \right),
\end{equation}
and
\begin{eqnarray} \sla{pe6}
\ddot{\delta \phi} &+& 3H\dot{\delta \phi} + \left[ \frac{k^2}{a^2} + m_{\phi}^2 + \epsilon \rho \right] \delta \phi \nonumber\\
&=& - 2  m_{\phi}^2 \phi_0 \Psi - \epsilon \rho_0 \phi_0 (\delta + 2 \Psi) + 4 \dot{\phi}_0 \dot{\Psi}.
\end{eqnarray}

These six equations are not all independent, we use (\ref{pe2}), (\ref{pe4}), (\ref{pe5}) and (\ref{pe6})  to numerically evolve a set of initial 
perturbations. The background dynamics are the same as in figure \ref{fig2}. In fact, for the case $\beta =0.01$, they yield the best fit to 
supernovae data  in the last section. The results of the numerical analysis are shown in figures
\ref{fig5} and \ref{fig6}. The dashed lines are the results in the $\Lambda$CDM model. 

An interesting parameter is the squared of the effective
sound speed of the perturbations of the scalar field, $c_s^2 = \delta p_{\phi} / \delta \rho_{\phi}$,  
\begin{equation} \sla{ss}
c^2_s = \frac{\dot{\phi}_0 \dot{\delta \phi} - \dot{\phi}_0^{\,2}\Psi - m_{\phi}^2\phi_0 \delta \phi }{\dot{\phi}_0 \dot{\delta \phi} - 
\dot{\phi}_0^{\,2}\Psi + m_{\phi}^2\phi_0 \delta \phi + \frac{1}{2} \epsilon \phi_0^2(\delta + \frac{\epsilon \phi_0 }{1+ 
\frac{1}{2} \epsilon \phi_0^{\,2}}\delta\phi)} .
\end{equation}

For an uncoupled scalar field dark energy component, as quintessence, the sound speed is exactly one in the slow-roll 
approximation, preventing scalar field structure to form. In our model, the 
scalar field acts both as dark energy and, by the coupling to baryons, as dark matter. 
As a dark energy component,  scalar field perturbations tend to be erased, but on the other hand the interaction drives to zero the 
speed of sound squared, due to the last term in the denominator of equation (\ref{ss}), allowing the scalar field perturbations to grow. As 
a result of this interplay scalar field perturbations 
are not completeley damped, instead initial scalar field inhomogeneities could be amplified or at least frozen. This behavior is shown in figures 
\ref{fig5} and \ref{fig6}, where the scalar field perturbations do not decay to zero as the time goes on, but they tend to a negative value about 
which they oscillate. 

The scalar field perturbations become a source of  the gravitational potential (see equation \ref{pe1}) that acts upon the baryonic matter 
through equation (\ref{pe4}). This is a feature of dark matter.  However, note that the sign of the average of the scalar field perturbations 
is negative and, from the Poisson equation (46), they act as a repulsive gravitational source. Nevertheless, the gravitational potential, read 
from figures \ref{fig5} and \ref{fig6}, is  similar to the one in the $\Lambda$CDM model. This is because the baryonic matter energy density 
appears multiplied by the factor $(1+\epsilon \phi^{2}/2)$  in equation (\ref{pe1}) and it contributes  as an attractive gravitational source.  
On the other hand, the scale factor grows at slower rate 
than in the $\Lambda$CDM  model and this yields an increase in the gravitational perturbations through the second term of 
the left-hand side. of equation (\ref{pe1}).  As a net result of these effects the gravitational potential is slightly stronger ($\Psi$ more negative), 
and therefore the density contrast grows faster, than in the $\Lambda$CDM model. These can be seen in figures \ref{fig5} and \ref{fig6}.
 
 As it was pointed out in  \cite{Bean08_2} coupled dark energy models could suffer from unwanted severe instabilities characterized by a 
 negative speed of sound squared. This is not the case  in our model  in which the numerical analysis shows that the scalar field perturbations 
 do not grow but instead oscillate around a nonzero value. A more detailed analysis \cite{Corasaniti08}  has shown that these 
 instabilities  are suppressed when the scalar field is in the slow-roll regime, as in our case. 

\section{Conclusions}

It is possible that future observations and experiments reveal that there exist some interactions between the dark sector and the known fields of the 
standard model; indeed, this is the hope for terrestrial experiments in detecting dark matter. The dark degeneracy allows
us to study a wide variety of models that mimic the $\Lambda$CDM at today's observation accuracy but with a richer dynamics. 

In this paper we have presented a plausible mechanism in which an interaction between a quintessence-like field and baryonic matter gives rise to
effects similar to those of dark matter. The interaction is given through the trace of the energy momentum tensor of the matter fields, 
and in the case that we have considered, dust matter, 
it is the same as an interaction of the type $\,f(\phi)\mathcal{L}$. The differences are manifest if we instead use a radiation field, {\it e.g.} 
photons, in this case the interaction vanishes -neglecting the trace anomaly \cite{BirrelDavies}- and in this sense is dark. Because of this coupling the 
radiation era is the same as in the standard big bang and the differences appear once matter dominates. 
We have described the background cosmology and by fitting the parameters of the model we can mimic as far as we want the late
time  $\Lambda$CDM background solutions, from decoupling until today. Specifically we have shown the differences for the cases of $\beta = 0.4$ 
($\epsilon \approx G$) and  $\beta=0.01$. For this last value of $\beta$, we made fits to the Union 2 supernovae data set and 
found that the best fit is obtained by choosing the other two parameters of the model equal to $C_1=0.305\,$ and $\,C_2(\phi_0) = 4.66$.

We have worked out the first order perturbation theory and shown that these models are capable to form linear structure 
without dark matter. Initial perturbations of a free scalar field dark energy are erased 
because its sound speed is equal to one,  
and by the fact that dark energy accelerates the Universe, it tends to freeze any matter perturbation. One of the effects of the 
interaction, when the field is slow-rolling, is to decrease the sound speed. In our model this induces 
the scalar field perturbations to oscillate about a nonzero negative value and yield a repulsive gravitational force over the baryonic matter 
perturbations. Despite this effect, we showed that the baryonic matter density contrast  could grow as fast as in the  $\Lambda$CDM model. 
This is because  the interaction enhances the gravitational strength of the baryons.  This latter effect increases as $\beta$ does, 
but at the same time the increase of $\beta$ can break the slow-roll. Thus, suitable $\beta$ are found for  $\beta \lesssim 0.04$.

\section{Appendix}

In this Appendix, we show how the interaction affects  different types of matter, namely a fermion field and  a perfect fluid.

For a fermion Dirac field we focus on the microscopic interaction and we consider a fixed Minkowski spacetime. The Lagrangian is 
$\mathcal{L} = \mathcal{L}_{\psi} +  \mathcal{L}_{int} + \mathcal{L}_{\phi} = (-\bar{\psi}( i \eta^{\mu\nu} \gamma_{\mu}\partial_{\nu} - m)\psi + A(\phi) 
T)\sqrt{-\eta} + \mathcal{L}_{\phi}\,$, and the trace is given by equation (\ref{tracedef}), which now reads  
\begin{equation} \sla{tracedef2}
T = - \frac{2}{\sqrt{-\eta}} \frac{\delta (S_{\psi} + S_{int})}{\delta \eta^{\mu\nu}} \eta^{\mu\nu}.
\end{equation}
For a free fermion field, one obtains $\,T^{(0)}=-i\bar{\psi}\gamma_{\mu}\partial_{\mu}\psi = -m\bar{\psi}\psi\,$, where  we have
used the Dirac equation.
Now, for the total Lagrangian we make the ansatz $\,T = m e^{\alpha(\phi)} \bar{\psi}\psi\,$,  and inserting it into equation (\ref{tracedef2}), we obtain  
that $\,e^{\alpha(\phi)} = (1-A(\phi))^{-1}$. So, the Lagrangian now reads 
\begin{equation}
\mathcal{L} = -\bar{\psi}( i \gamma^{\mu}\partial_{\mu} - e^{\alpha(\phi)} m)\psi + \mathcal{L}_{\phi} .
\end{equation}
 
The effect of the interaction is to shift the mass of the fermions from $m$ to  $e^{\alpha(\phi)} m$.

Now, let us consider in the matter sector a perfect fluid with an equation of state parameter $w$ constant. The matter Lagrangian is 
$\mathcal{L}_m = - \rho \sqrt{-g}$, where $\rho$ is the energy density of the fluid in its rest frame \cite{deFelice}. 
We will assume that the trace of the energy momentum tensor depends only on $\rho$, $T=T(\rho)$.
It also depends on the metric, but only through $\rho\,$-. Then $\delta T = (dT/d\rho) \delta \rho\,$. Let us consider the action

\begin{equation}
S_{int} + S_{m} = \int d^4 x \sqrt{-g} ( A(\phi) T^{(m)} - \rho\,),
\end{equation}
performing variations with respect to the metric, equation (\ref{tracedef}) gives

\begin{eqnarray} \sla{EMT}
T_{\mu\nu} &=&   \rho(1+w)u_{\mu}u_{\nu} + w \rho \,g_{\mu\nu} + A(\phi) T g_{\mu\nu} \nonumber\\
           & &   - A(\phi) \frac{dT}{d\rho}  \rho (1+w) (u_{\mu}u_{\nu} + g_{\mu\nu}),
\end{eqnarray}
where we have used the identity $\,\delta \rho = \frac{1}{2} (\rho + p)(u_{\mu}u_{\nu} + g_{\mu\nu}) \delta g^{\mu\nu}$ (see for example \cite{deFelice}).
Taking the trace, we obtain the differential equation

\begin{equation}
3 A(\phi) (1 + w)  \rho \frac{ d T}{d \rho} + (1 - 4A(\phi)) T + (1-3w) \rho = 0,
\end{equation}
with general solution 

\begin{equation}
T = - \frac{(1-3w)}{1-(1-3w) A(\phi)}\, \rho + C \rho^{-(1-4 A)/3A(1+w)},
\end{equation}
where $C$ is an integration constant that we have to choose equal to zero if we want for a 
radiation fluid $T=0$. Then we obtain the next equation for the trace

 \begin{equation}
T = - \frac{1-3w}{1-(1-3w)A(\phi)} \rho.
\end{equation}
For dust, $w = 0$ and this equation reduces to (\ref{trace1}). For a relativistic fluid $w=1/3$ and the trace vanishes. 
This result agrees with the one obtained through an iterative procedure in \cite{SinghPad88,SamiPad03}.

\begin{acknowledgments}
J.L.C.C. thanks the Berkeley Center for Cosmological Physics for hospitality, and gratefully acknowledges support from a UC MEXUS-CONACYT Grant, and a CONACYT Grant No. 84133-F. A.A. acknowledges a CONACYT Grant No. 215819.

\end{acknowledgments}

\end{document}